# Resonance as a probe of the electron superconducting gap in BaFe$_{1.9}$Ni$_{0.1}$As$_2$

Jun Zhao[1], Louis-Pierre Regnault[2], Chenglin Zhang[1], Miaoying Wang[1], Zhengcai Li[3], Fang Zhou[3], Zhongxian Zhao[3], and Pengcheng Dai[1,3,4]

[1] Department of Physics and Astronomy, The University of Tennessee, Knoxville, Tennessee 37996-1200, USA

[2] Institut Nanosciences et cryogénie, SPSMS-MDN, CEA-Grenoble, F-38054 Grenoble Cedex 9, France

[3] Beijing National Laboratory for Condensed Matter Physics, Institute of Physics, Chinese Academy of Sciences, Beijing 100190, China

[4] Neutron Scattering Science Division, Oak Ridge National Laboratory, Oak Ridge, Tennessee 37831, USA

**The discovery of high-transition temperature (high-$T_c$) superconductivity near antiferromagnetism in iron arsenides raised the possibility of an unconventional superconducting mechansim[1-8]. The observation of clear Fermi surfaces and nodeless superconducting gaps by angle resolved photoemission[9-12] suggests that electron pairing in these materials may be mediated by quasiparticle excitations between sign reversed hole and electron Fermi pockets[5-8]. Although the presence of a "resonance" in the spin excitation spectrum found by inelastic neutron scattering[13-17] is consistent with this picture[18-20], there has been no direct evidence connecting the resonance to the superconducting gap energy. Here we show that for the optimally electron doped BaFe$_{1.9}$Ni$_{0.1}$As$_2$ ($T_c$ =20 K, Fig. 1c) iron arsenide**

**superconductor, application of a magnetic field that suppresses the superconductivity and superconducting gap energy also reduces the intensity and energy of the resonance. These results suggest that the energy of the resonance is proportional to the electron pairing energy, and thus indicate that spin fluctuations are intimately related to the mechanism of superconductivity in iron arsenides.**

Soon after the discovery of superconductivity in LaFeAsO$_{1-x}$F$_x$ (ref. 1), band-structure calculations of Fermi surfaces for these materials found two hole cylinders around the $\Gamma$ point and two electron cylinders around the M point[18]. In theories of spin fluctuation mediated superconductivity[5-8,18-21], electron pairing arises from quasiparticle excitations from the hole pocket to electron pocket (inset in Fig. 1c). While the normal-state spin excitations are dominated by the continuum of scattering, superconductivity arising from sign reversed interband scattering induces a resonance peak at the antiferromagnetic (AF) ordering wave vector $Q = (0.5, 0.5, 0)$ in the spin excitations spectrum (Fig. 1b). The energy of the resonance should be at (or slightly less than) the addition of hole and electron superconducting gap energies ($\hbar\omega = |\Delta(k + Q)| + |\Delta(k)|$) refs. 18-22. Although the resonance and its temperature dependence observed by inelastic neutron scattering in Ba$_{0.6}$K$_{0.4}$Fe$_2$As$_2$ (ref. 13), BaFe$_{2-x}$(Co,Ni)$_x$As$_2$ (refs. 14-17) are consistent with this picture, there has been no direct proof that the resonance energy is related to superconducting energy gap and therefore it is still unclear whether the mode is related to electron pairing. One way to resolve this problem is to study the effect of a magnetic field on superconductivity and spin excitations. A magnetic field can suppress $T_c$ and reduce the magnitude of the superconducting energy gap via either orbital pair breaking of Cooper pairs in the superconducting state or Pauli paramagnetism due to Zeeman effect on electron spins. If the resonance is associated with quasiparticle excitations across the electron and hole pockets[5-8], application of a magnetic field that suppresses the superconducting gaps should also reduce the energy of the resonance. We find this is indeed the case for BaFe$_{1.9}$Ni$_{0.1}$As$_2$ (Fig. 1), and our results thus provide the most compelling evidence that electron pairing in iron arsenide superconductors is directly correlated with magnetic excitations.

In the undoped state, the parent compounds of iron arsenide superconductors are nonsuperconducting antiferromagnets with a spin structure as shown Fig. 1a (refs. 3,4). Upon doping to induce optimal superconductivity, the static AF order is suppressed and low-energy magnetic excitations in the superconducting state are dominated by a spin gap and resonance above the spin gap energy[13-17]. For optimally electron-doped superconductor BaFe$_{1.9}$Ni$_{0.1}$As$_2$ with $T_c = 20$ K (Fig. 1c), the resonance occurs near $\hbar\omega \approx 8$ meV at $Q = (0.5,0.5,0)$ reciprocal lattice unit (rlu) above a $\hbar\omega \approx 3$ meV spin gap in the low temperature superconducting state[15,16]. We used inelastic neutron scattering to study the effect of a 14.5-Tesla $c$-axis aligned magnetic field on the resonance and spin gap (Fig. 1). At zero field, energy scans in the normal state ($T = 25$ K) show clear gapless continuum of scattering at the signal $Q = (0.5,0.5,0)$ position above the background $Q = (0.62,0.62,0)$ (red filled and open circles in Fig. 1d). On cooling into the superconducting state ($T = 2$ K), a spin gap opens below $\hbar\omega \approx 3$ meV and the low energy spectral weight is transferred into the resonance at $\hbar\omega \approx 8$ meV (refs. 15,16). While imposition of a 14.5-T magnetic field along the $c$-axis has little effect on the background (Fig. 1d) and normal state scattering at $Q = (0.5,0.5,0)$ (not shown for the purpose of clarity), the resonance peak in the superconducting state is clearly suppressed and shifted to a lower energy (blue triangles in Fig. 1d). Figure 1e plots the temperature dependence of the imaginary part of the dynamic susceptibility $\chi''(Q,\omega)$, obtained by subtracting the background scattering and correcting for the Bose population factor $\chi''(Q,\omega) = [1-\exp(-\hbar\omega/(k_B T))]S(Q,\omega)$, where $k_B$ is the Boltzmann constant. Inspection of the Figure reveals that application of a 14.5-T magnetic field shifted the energy of the resonance from $\hbar\omega \approx 7.8 \pm 0.15$ meV to $6.5 \pm 0.2$ meV, and broadened the mode only slightly. Comparison of the temperature difference plots at zero and 14.5-T in Fig. 1f confirms the shift in energy of the mode. In addition, the data suggest that superconductivity-induced resonance intensity gain (the shaded area in Fig. 1f for zero field) decreases about 23% from zero to 14.5-T.

Although the constant-$Q$ scans in Fig. 1 are excellent ways of determining the influence of a magnetic field on the resonance mode energy and peak intensity, they do not provide information on how the field affects the momentum distribution of the magnetic excitations (spin-spin correlations). Figure 2 summarizes $Q$-scans at energies $\hbar\omega = 0, 2, 3, 8$ meV which corresponds to elastic scattering, below and near spin gap energy, and at the resonance energy, respectively. At zero energy transfer ($\hbar\omega = 0$ meV) and 2 K, the scattering across $Q = (0.5, 0.5, 0)$ are featureless at zero and 14.5-T (Fig. 2a), indicating that such a field does not induce AF long range static order. For $\hbar\omega = 2$ meV, the scattering at zero field show no peak, which is consistent with the presence of a spin gap at 2 K (refs. 15,16). However, the identical $Q$-scan at 14.5-T shows a clear peak at $Q = (0.5, 0.5, 0)$, suggesting a field-induced scattering due to the decreasing value of the zero field spin gap (Figs. 1e and 2b). Similarly, a 14.5-T field enhances the zero field $\hbar\omega = 3$ meV peak near $Q = (0.5, 0.5, 0)$ in the superconducting state at 2 K (Fig. 2c) but has no effect above $T_c$ at 25 K (Fig. 2e). In contrast, imposition of a 14.5-T field at 2 K suppresses the resonance intensity at $\hbar\omega = 8$ meV (Fig. 2d). The same field again has no effect in the normal state at 25 K (Fig. 2f). Fourier transforms of the Gaussian peaks at $\hbar\omega = 8$ meV and 2 K in Fig. 2d give spin-spin correlation lengths of $\xi = 57 \pm 2$ Å and $\xi = 53 \pm 3$ Å for 0 and 14.5-T, respectively. These values are larger than the superconducting coherence length of $27.6 \pm 2.9$ Å, but much smaller than the zero temperature London penetration depth of $\lambda(0) \approx 2000$ Å determined for $BaFe_{2-x}Co_xAs_2$ with $T_c$'s ~ 22-25 K (refs. 24,25). Whereas a field can change the energy and intensity of the resonance, it has small effect on spin-spin correlation length. We note that magnetic field also has no effect on spin-spin correlation length for copper oxide superconductor $YBa_2Cu_3O_{6.6}$ (ref. 23).

Figure 3 compares temperature dependence of the scattering at $Q = (0.5,0.5,0)$ for $\hbar\omega = 2$ and 8 meV at zero and 14.5-T, respectively. Consistent with previous work[15,16], we find that a spin gap opens at $\hbar\omega = 2$ meV (Fig. 3a), and the scattering at the resonance energy ($\hbar\omega = 8$ meV) shows a superconducting order parameter-like increase below $T_c$ (Fig. 3c). Upon application of a 14.5-T field, the kink in the zero field temperature dependence data at $\hbar\omega = 2$ meV indicative of the opening of the spin gap disappears (Fig. 3b). Instead, the scattering shows no observable anomaly in the probed temperature range. On the other hand, temperature dependence of the scattering at the resonance energy ($\hbar\omega = 8$ meV) shows a clearly depressed $T_c$ of ~16 K at 14.5-T from $T_c = 20$ K at zero field (Figs. 3c and 3d). Since an applied magnetic field that suppresses $T_c$ also decreases the superconducting gap energy, these results demonstrate that the resonance energy and its temperature dependence are directly correlated with the superconducting gap energy and electron pairing strength.

Figures 4a and 4b show the magnetic field dependence of the scattering at the resonance energy in the superconducting state at 2 K and normal state at 25 K, respectively. While the normal state spin excitations have no observable field effect up to 14.5-T (Fig. 4b), the scattering at the resonance energy clearly decreases with increasing field (Fig. 4a). The solid line is a linear fit to the data using $I/I_0 = 1 - B/B_{char}$ with $B_{char} \approx 32$ T, where intensity of the resonance is suppressed to the normal state value. The dotted line represents a fit assuming $I/I_0 = 1 - (B/B_{char})^{1/2}$, where $B_{char} \approx 66$ T (ref. 23). Since the energy of the resonance is decreasing with increasing field, it is difficult to estimate the characteristic field $B_{char}$ using the field dependent scattering at the resonance energy at zero field and compare with the upper critical field $B_{c2}$. We note, however, that scanning tunneling spectroscopy and magnetotransport measurements on BaFe$_{1.8}$Co$_{0.2}$As$_2$ samples ($T_c \approx 22 - 25.3$ K) showed an upper critical field of ~43-T (ref. 24) and ~50-T (ref. 26), respectively, for a $c$-axis aligned field.

The total momentum sum rule states that the magnetic structure factor $S(Q, \omega)$, when integrated over all wavevectors and energies, $i.e.$, $\int_{-\infty}^{\infty} d\omega \int dQ S(Q, \omega)$, should be a temperature- and field-independent constant[27]. To see if this is true at zero and 14.5-T, we plot in Figure 4c experimentally measured difference spectrum, $S(Q, \omega, B = 0 \text{ T}) - S(Q, \omega, B = 14.5 \text{ T})$, at $Q = (0.5, 0.5, 0)$ and 2 K. We find that the spectral weight loss of the resonance under a 14.5-T field is approximately compensated by the field-induced subgap intensity gain, suggesting that the sum rule is satisfied within our probed $Q$-energy space.

In previous work on copper oxide superconductors, application of a magnetic field was found to suppress the intensity of the resonance[23] and induce AF order at the expense of the resonance[28-32]. However, the energy of the resonance was not found to change with field[23,28-32]. Theoretically, several effects of a magnetic field on the resonance and spin excitations have been considered within the random-phase approximation[33]: first, the supercurrents circulating around the field-induced vortices may broaden the resonance in energy without changing its $Q$-energy integrated weight; second, a field-induced uniform suppression of the superconducting gap magnitude should cause the resonance to shift to lower energy and decrease in intensity; third, the effect of field-induced suppression of the superconducting coherence factor might lead to suppression of the spectral weight and causing the resonance to shift to higher energy; and finally, suppression of the resonance within the field-induced vortex cores could result in reduced resonance intensity without shifting its position, consistent with neutron scattering results on $YBa_2Cu_3O_{6.6}$ (refs. 23,33). Since we observed clear field-induced resonance energy and intensity reduction in $BaFe_{1.9}Ni_{0.1}As_2$ (Figs. 1-4), our data are most consistent with a field-induced suppression of the superconducting gap energy.

If this microscopic picture is indeed correct, we can use neutron scattering data in Figs. 1-4 to estimate the upper critical field $B_{c2}$ and expected resonance energy shift at 14.5-T field. In Ginzburg-Landau theory, magnetic field dependence of the superconducting gap $\Delta(B)$ is related to the zero field gap $\Delta(0)$ via $\Delta(B)/\Delta(0) = \sqrt{1 - B/B_{c2}}$ (ref. 33). Since superconducting gap is proportional to $T_c$ (*i.e.* $2\Delta \propto k_B T_c$, refs. 9,11), we estimate $B_{c2} = 40.3$-T using the measured $T_c$ ($\approx 16$ K) at 14.5-T in Fig. 3d and $B_{c2} = B/[1 - (T_c(14.5 \text{ T})/T_c(0 \text{ T}))^2]$. This value is very close to the measured $B_{c2} = 43 - 50$ T for BaFe$_{1.8}$Co$_{0.2}$As$_2$ which have slightly higher $T_c$'s (refs. 24,26). If the resonance energy is associated with the superconducting gap energy via $\hbar\omega = |\Delta(k+Q)| + |\Delta(k)|$, one should expect the resonance energy to shift from $\hbar\omega \approx 7.8 \pm 0.15$ meV at zero field to $\hbar\omega(14.5\text{ T}) = (T_c(14.5\text{ T})/T_c(0\text{ T}))\hbar\omega(0\text{ T}) \approx 6.24$ meV. Inspection of Fig. 1e shows that this is indeed the case with experimental observation of $\hbar\omega(14.5\text{ T}) = 6.5 \pm 0.2$ meV. This is the most compelling evidence that the resonance is related to superconducting gap energy. Although our observation of a field-induced resonance intensity reduction is also consistent with field-suppressed superconducting gap picture[33], the multiband nature of the system[5-8] means that one needs a more detailed theoretical calculation to compare with the experiments.

Finally, to test if the resonance directly probes the electron spin singlet-to-triplet transition (from singlet spin $S = 0$ for Cooper pairs to triplet spin $S = 1$), we note that the Zeeman magnetic energy for a 14.5-T field is at $\pm g\mu_B B \approx \pm 1.7$ meV (assuming the Lande factor $g = 2$ and $S = 1$). Experimentally, the energy widths of the resonance in Fig. 1e change from $4.2 \pm 0.34$ meV full-width-at-half maximum (FHWM) at zero field to $4.7 \pm 0.53$ meV FHWM at 14.5-T. Given the finite energy width of the resonance and instrumental resolution (Fig. 1e), we find no conclusive evidence for the Zeeman splitting of the resonance. Therefore, while our data support the notion that the resonance is directly correlated with the superconducting electron energy gap, it remains unknow whether the mode is the long-sought singlet-to-triplet transition. Regardless whether this is the case, our data suggest that magnetic excitations are the most promising candidate for mediating the electron pairing for superconductivity in iron arsenides.

**Acknowledgements** We thank T. A. Maier, M. R. Norman, Y. Yin, and Jiangping Hu for helpful discussions, and Oliver Lipscombe for a critical reading of the manuscript. This work is supported by the US NSF and DOE Division of Materials Science, Basic Energy Sciences. This work is also supported by the US DOE through UT/Battelle LLC. The work at IOP is supported by the Chinese Academy of Sciences and the Ministry of Science and Technology of China.

**Author Information** The authors declare no competing financial interests. Correspondence and requests for materials should be addressed to P.D. (daip@ornl.gov).

**Figure 1 Magnetic structure, probed reciprocal lattice space and magnetic field dependence of the scattering at the AF wavevector for BaFe$_{1.9}$Ni$_{0.1}$As$_2$.** Our inelastic neutron scattering experiments were carried out on the IN22 thermal triple-axis spectrometer at the Institut Laue-Langevin, Grenoble, France. We co-aligned 5.5 grams of single crystals of BaFe$_{1.9}$Ni$_{0.1}$As$_2$ grown by self-flux (with in-plane mosaic of 2 degrees). We define the wave vector $Q$ at $(q_x, q_y, q_z)$ as $(H, K, L)$ = $(q_x a/2\pi, q_y b/2\pi, q_z c/2\pi)$ in reciprocal lattice units (rlu), where $a = b = 3.963$, and $c = 12.77$ Å are the tetragonal unit cell lattice parameters (refs. 15,16). Our samples are aligned in the $(H,K,0)$ horizontal scattering plane inside a 14.5-T vertical field magnet. The final neutron energy was fixed at 14.7 meV with a pyrolytic graphite filter before the analyzer. Field was always applied in the normal state at 25 K. a) Schematic diagram of the Fe spin ordering in BaFe$_2$As$_2$. b) Reciprocal space probed and the direction of applied field. c) Susceptibility of our sample indicating $T_c = 20$ K. The inset shows schematic diagram of how the resonance is produced by quasiparticle excitations between the hole and electron pockets. d) Energy scans at the signal $Q = (0.5, 0.5, 0)$ and background $Q = (0.62, 0.62, 0)$ rlu positions for various fields and temperatures. The background scattering has negligible temperature and field dependence. e) Temperature and field dependence of $\chi''(Q,\omega)$ at $Q = (0.5, 0.5, 0)$. Horizontal bar indicates instrumental energy resolution. f) Difference spectra of the neutron intensity between $T = 2$ K ($< T_c$) and $T = 25$ K ($T_c+5$ K) at $Q = (0.5, 0.5, 0)$ for $B = 0$ and a 14.5-T c-axis aligned field. Error bars indicate one sigma.

**Figure 2 A series of constant-energy (H,H,0) scans through the AF wavevector Q = (0.5,0.5,0) as a function of increasing energy at different temperatures and fields.** a) $\hbar\omega = 0$ meV; b) $\hbar\omega = 2$ meV. Spin-spin correlation length at 2 K and 14.5-T is $\xi = 64 \pm 16$ Å. Note that the vertical scales for the $B = 0$-T data in a) and b) were offset for clarity; c) $\hbar\omega = 3$ meV at 2 K. At zero field, $\xi = 65 \pm 10$ Å. At 14.5-T, $\xi = 47 \pm 10$ Å; d) $\hbar\omega = 8$ meV at 2 K. At zero field, $\xi = 57 \pm 2$ Å. At 14.5-T, $\xi = 53 \pm 3$ Å; e) $\hbar\omega = 3$ meV at 25 K. At zero field, $\xi = 62 \pm 5$ Å. At 14.5-T, $\xi = 54 \pm 6$ Å; f) $\hbar\omega = 8$ meV at 25 K. At zero field, $\xi = 55 \pm 5$ Å. At 14.5-T, $\xi = 49 \pm 4$ Å. The solid lines are Gaussian fits to the data on linear backgrounds and horizontal bars in b)-f) are the instrumental resolution. Error bars indicate one sigma.

**Figure 3 Effect of a magnetic field on the temperature dependence of the resonance and low-energy spin excitations at Q = (0.5,0.5,0).** a) Temperature dependence of the scattering at $\hbar\omega = 2$ meV and zero field shows the opening of a spin gap slightly below $T_c$ (refs. 15,16). b) The same temperature dependence at 14.5-T. The kink is now gone. c) Temperature dependence of the scattering at the resonance energy of $\hbar\omega = 8$ meV and zero field displays order parameter like intensity increase below $T_c = 20$ K. d) Application of a 14.5-T field suppresses $T_c$ to ~16 K. Error bars indicate one sigma.

**Figure 4 Magnetic field dependence of the resonance below and above $T_c$ and test of the total moment sum rule.** a) The magnetic field dependence of the scattering at $\hbar\omega = 8$ meV, $Q = (0.5,0.5,0)$, and 2 K. While the solid line is a fit using $I/I_0 = 1 - B/B_{char}$ with $B_{char} \approx 32$ T, the dotted line represents $I/I_0 = 1 - (B/B_{char})^{1/2}$, where $B_{char} \approx 66$ T. b) The scattering at 25 K has no observable field dependence. c) The difference spectrum of the neutron scattering intensities between zero and 14.5-T field at 2 K and $Q = (0.5,0.5,0)$. The scattering should be centered around zero if spin excitations are not affected by the field. Positive scattering indicate field-induced suppression while negative scattering represents field-induced enhancement.

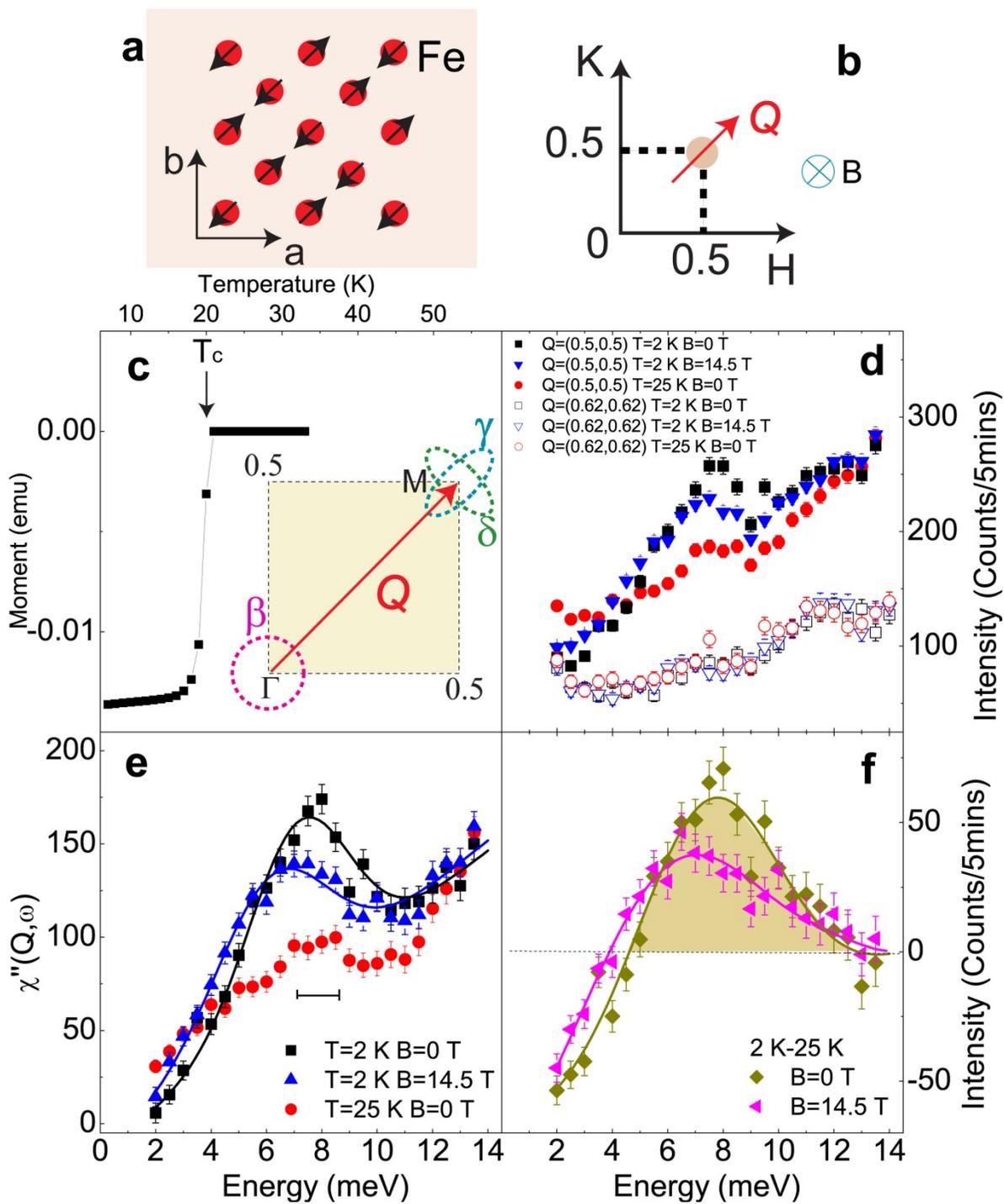

Fig. 1

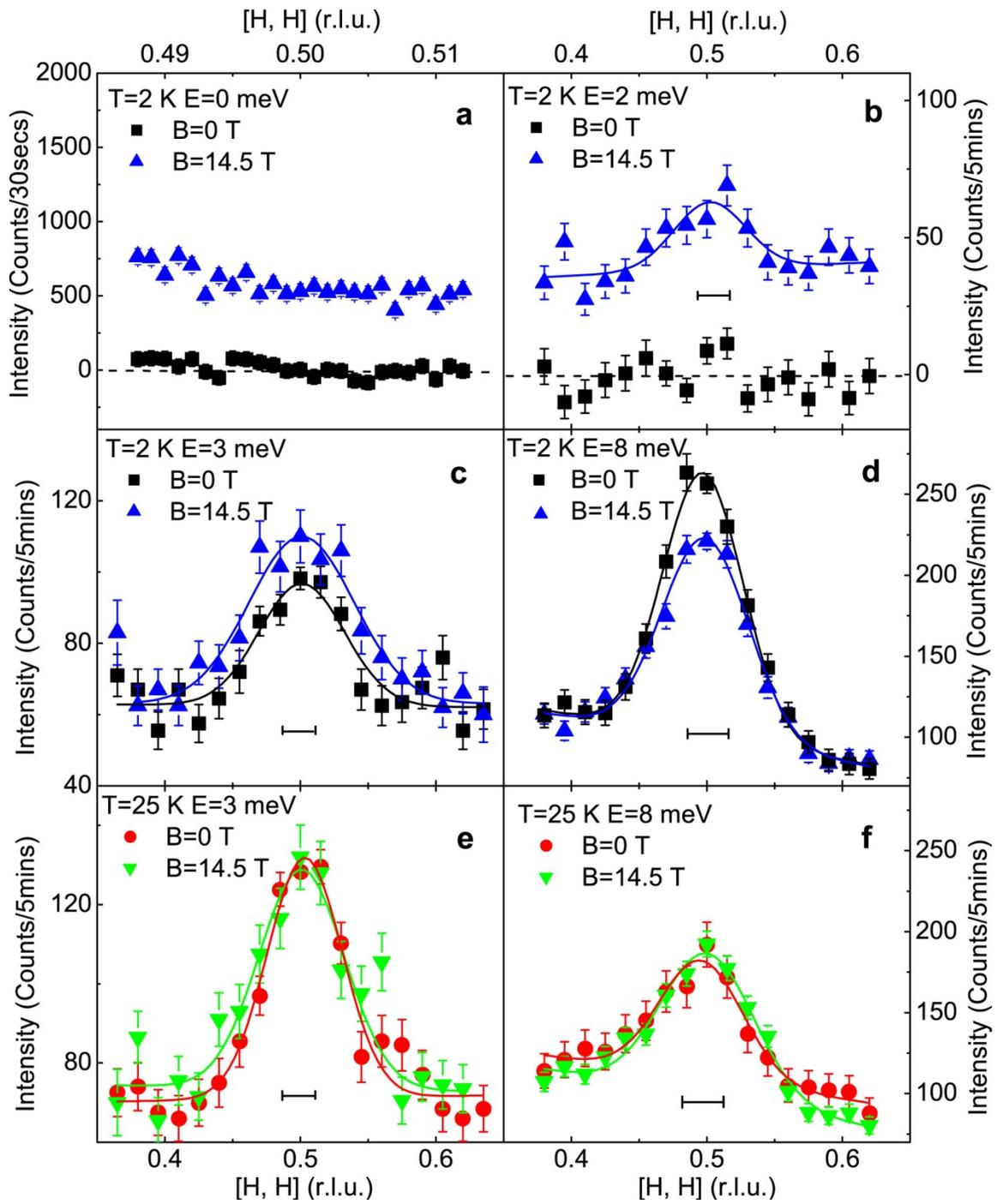
Fig. 2

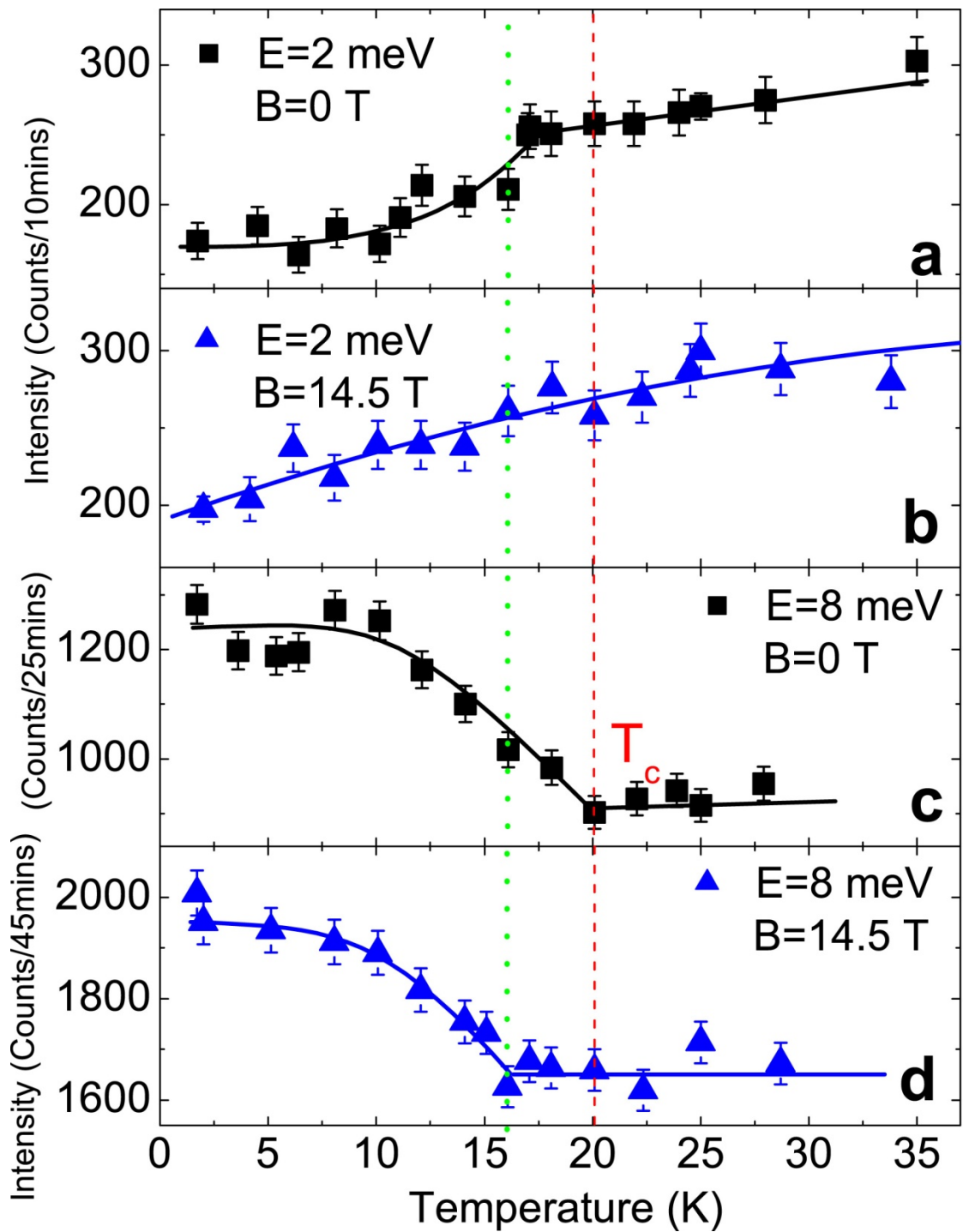

Fig. 3

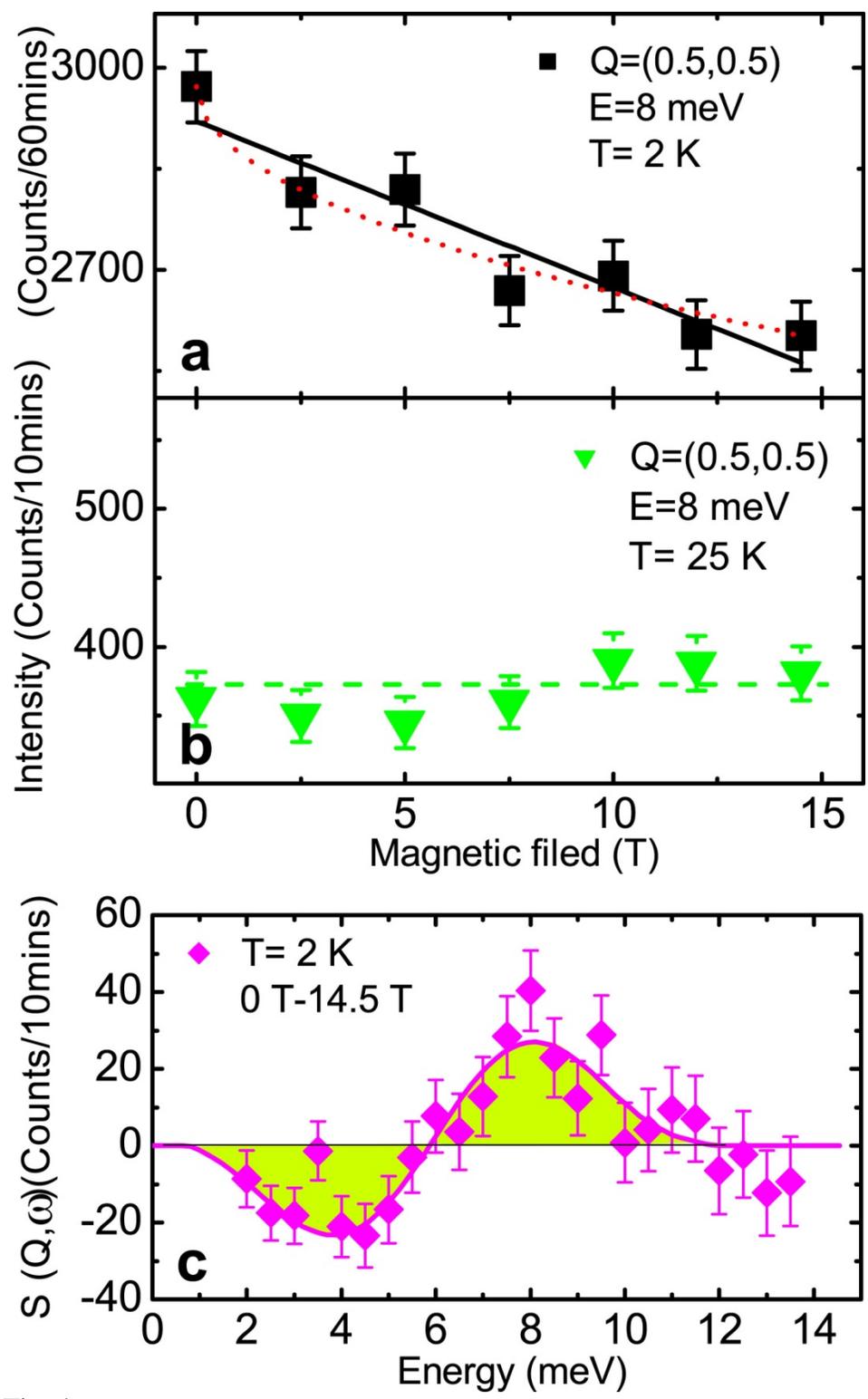

Fig. 4